\documentclass{iopart}
\usepackage{epsfig}
\begin{document}
\title{Single-impulse magnetic focusing of launched cold atoms}
\author{Matthew J Pritchard\dag, Aidan S Arnold\ddag, David A Smith\dag~and Ifan G Hughes\dag}
\address{\dag~Department of Physics, Rochester Building, University of
Durham, South Road, Durham, DH1 3LE, UK}
\address{\ddag~Department of Physics, University of Strathclyde, Glasgow, G4 0NG, UK}
\ead{i.g.hughes@durham.ac.uk}
\date{\today}

\begin{abstract}
We have theoretically investigated the focusing of a launched cloud of cold atoms.
Time-dependent spatially-varying magnetic fields are used to impart impulses leading to a
three-dimensional focus of the launched cloud. We discuss possible coil arrangements for
a new focusing regime: isotropic 3D focusing of atoms with a single magnetic lens. We
investigate focusing aberrations and find that, for typical experimental parameters, the
widely used assumption of a purely harmonic lens is often inaccurate.
The baseball lens offers the best possibility for
isotropically focusing a cloud of weak-field-seeking atoms in 3D.
\end{abstract}

\section{Introduction}

The development of laser-cooling techniques~\cite{Adams97} has facilitated the
preparation of samples of atoms at microKelvin temperatures~\cite{nobel97}. It is
therefore now possible to modify drastically  the centre-of-mass motion of
atoms, in direct contrast with the small angular deflection of fast beams
studied prior to the development of laser cooling~\cite{Ramsey}. The (very
strong) optical scattering force utilised to cool atoms can also be used to manipulate
their position, as can the (relatively weak) conservative optical dipole force. Since the
dipole force is coherent it represents a better way of manipulating
cooled atoms than the scattering force (for which the random nature of
absorption and spontaneous emission leads to Brownian heating). An alternative
method of manipulating paramagnetic cold atoms is to use the
{Stern-Gerlach force~\cite{Hinds}.

One of the goals in the field of atom optics~\cite{Adams94} is to
realise atom-optical elements that are analogues of conventional
optical devices, such as mirrors, lenses and beam-splitters. An atom
mirror reverses the component of velocity perpendicular to the
surface and maintains the component parallel to the surface. An atom lens can modify
both the transverse velocity component and the longitudinal component. To date,
the Stern-Gerlach force has been
used to realise flat atomic mirrors~\cite{flat}, curved atomic
mirrors~\cite{curved}, and pulsed mirrors for both cold
(thermal)~\cite{kadio} and Bose condensed atoms~\cite{BECmirror}. It
has also been demonstrated that the surface of a magnetic mirror can
be adapted in real time with  corrugations that can be manipulated
in times shorter than the atom-mirror interaction
time~\cite{corrugated}.

There are many reasons for studying focusing of launched cold atoms: atom
lithography~\cite{meschede}; transferring cold atoms from a MOT to a remote vacuum
chamber of lower background pressure~\cite{Szymaniek99}; loading miniature magnetic
guides~\cite{minmagguide}, atom chips~\cite{atomchips} and storage rings
\cite{rings}. In comparison to an unfocused cloud, the density of the cloud
can be increased by many orders of magnitude after magnetic focusing.

The first demonstration of 3D focusing using pulsed magnetic lenses was
conducted by Cornell~{\it et al.}~\cite{mon1}. However, their work did not address the
optimum strategy for achieving a compact focused cloud, nor did they discuss the limiting
features for the quality of their atom-optical elements. The group of Gorceix has
performed experiments demonstrating the longitudinal Stern-Gerlach effect with an atomic
cloud using pulsed magnetic forces \cite{Marec}, and an experimental and theoretical
study of cold atom imaging by means of magnetic forces \cite{Gor}. To date, the
experimental and theoretical studies of pulsed magnetic focusing have been analysed
under the assumption that the magnetic lens potential is harmonic - this work
addresses the validity of this approximation, and the effects of aberrations. We
note also that there has been no theoretical or experimental work on full 3D focusing of
cold atoms using a single magnetic pulse, although this is a very useful focusing
strategy.

The scope of this paper is to investigate theoretically and numerically  the
limiting factors to the quality and size of the final image obtained in pulsed magnetic
focusing experiments; to identify the sources  of aberration; and
to discuss schemes for minimising their deleterious effect.
In this paper we restrict our attention to focusing strategies
using a single magnetic impulse.  A second paper describing more than one impulse
(alternate gradient focusing)  \cite{matt2} is in preparation.
Whilst this work shall concentrate on the
analysis for achieving a compact cloud in space, it is also possible to use pulsed
magnetic fields to reduce the momentum spread of an expanding cloud with appropriate
magnetic impulses. This can be viewed as an implementation of $\delta$-kick cooling, which
has been demonstrated with atoms \cite{Myrsk}, ions \cite{Goldberg} and
Bose Einstein condensates (BEC)
\cite{BECmirror}. Some of the techniques described here are also successfully
used for the deceleration of polar molecules using time-varying electric fields
\cite{Meijer}. Atom-optical elements realised with light forces \cite{Adams97, Dowling96}
are beyond the scope of this paper.

The remainder of the paper is organised as follows: Section 2 outlines the theory of
how to achieve the desired magnetic fields;  Section 3 contains an analysis
of magnetic imaging and minimising the final cloud size; Section 4 describes and
contrasts the performance of different magnetic lenses;
Section 5 contains a discussion and concluding remarks.

\section{Magnetic lens theory}

\subsection{The Stern-Gerlach force}

An atom's magnetic dipole interaction energy is $U=m_{F} g_{F} \mu_{\rm B}B$ for a field
of magnitude $B,$ where $m_{F}$ is the magnetic quantum number, $g_{F}$ is the Land\'{e}
g-factor and $\mu_{\rm B}$ the Bohr magneton. We assume that field zeros are avoided and
that the magnetic moment adiabatically follows the field. Depending on an atom's
hyperfine quantum state, the Stern-Gerlach force $F=-\nabla U\propto\nabla B$ can be used
to attract it towards weak (if $m_F g_F>0)$ or strong (if $m_F g_F<0)$ magnetic fields.
The choice of whether atoms in weak or strong-field seeking states are launched depends
on the particular application. For some applications, e.g. loading a remote dipole trap,
or a secondary magneto-optical trap, it does not matter which atomic state is used. In
this work we assume that the ensemble (of alkali metal atoms) has been optically pumped
into the stretched state $\vert F=I+1/2, m_{F}=F \rangle$ (a weak-field seeking state
with $m_{F} g_{F}=1).$

A purely harmonic magnetic field magnitude will result in an aberration-free lens. In
this paper we will consider current-carrying wires assembled to give two different kinds
of second order magnetic field magnitude:
\begin{equation}
B_{1\rm{D}}(x,y,z)=B_{0}+\frac{B_2}{2}\left(-x^2/2-y^2/2+(z-z_{\rm c})^2\right),
\label{1D}
\end{equation}
\begin{equation}
B_{3\rm{D}}(x,y,z)=B_{0}+B_1 (z-z_{\rm c})+\frac{B_2}{2}\left(x^2 + y^2 + (z-z_{\rm
c})^2\right). \label{isot}
\end{equation}
$B_{0}$, $B_{1}$ and $B_{2}$ are the bias field, the axial gradient and the field curvature,
respectively. In both
cases these parameters are chosen to prevent the field magnitude exhibiting zeros in the
region of interest, thus avoiding Majorana spin-flip transitions \cite{petr}.  The point
$\{0,0,z_{\rm c}\}$ defines the centre of the lens. A lens of the form of $B_{1\rm{D}}$
can be used either to focus axially or radially;  a lens of the form of $B_{3\rm{D}}$ is
used to  focus isotropically in 3D. The accelerations associated with these fields are
harmonic about $\{0,0,z_{\rm c}\}:$
\begin{equation}
\textbf{a}_{1\rm{D}}= -{\omega}^2\{-x/2,-y/2,(z-z_{c})\},
 \label{force1D}
\end{equation}
\begin{equation}
\textbf{a}_{3\rm{D}}= -{\omega}^2\{x,y,(z-z_{c})\}+\textbf{a}_{0},
 \label{force}
\end{equation}
where ${\omega}^2=\mu_{\rm B}B_{2}/m$ is a useful measure of the power of the lens,
$\textbf{a}_{0}=\{0,0,\mu_{\rm B}B_{1}/m\}$ is a constant acceleration arising from axial
magnetic gradients, and $m$ is the atomic mass. Note that for the 1D case
(equation~(\ref{force1D}))
the axial curvature is twice the magnitude of and opposite in sign to the radial
curvature, ${\omega_\mathrm{r}}^2=-2 {\omega_\mathrm{z}}^2.$

The acceleration of equation (\ref{force1D}) results in a lens which is axially
converging (diverging) and radially diverging (converging) if $B_2$ is positive
(negative). In order to achieve a 3D focus with such lenses, an axially converging lens
pulse must be followed by an appropriately timed axially diverging lens (or vice versa).
This alternate-gradient focusing strategy is the subject of another paper~\cite{matt2}.
In contrast,  this paper details  novel lenses which yield accelerations of the form in
equation (\ref{force}).  Isotropic 3D focusing can be achieved with a single lens pulse
for atoms in weak-field-seeking states, as Earnshaw's theorem (which states that a
maximum of magnetic field magnitude in free space is not allowed \cite{earnshaw}) ensures
$B_2\geq 0$ in this case. Note that the harmonic accelerations in both
equation~(\ref{force1D}) and (\ref{force}) lead to three separable one-dimensional simple
harmonic oscillator equations for the atomic motion.

\subsection{Magnetic fields from current bars and circular  coils}
\label{circcoil}

This work considers  straight current bars and circular coils for the formation of
lenses. The Biot-Savart law yields magnetic fields that are analytic for both finite- and
infinite-length current bars. For circular coils the field can be expressed in terms of
elliptic integrals~\cite{smythe}. A discussion of the form of the contours of magnetic
field magnitude for various magnetic trapping configurations has been provided by
Bergeman~{\it et al.}~\cite{metcalf}. The fields are constrained by Maxwell's equations,
which, in conjunction with symmetry arguments, allow the spatial dependence of the fields
to be parameterised with a small number of terms. In particular, for a cylindrically
symmetric magnetic coil configuration, the fourth-order on-axis 1D magnetic field,
$B_z(r=0,z)=\sum_{i=0}^{4}{B_i\,z^i/i!},$ gives the complete fourth-order 3D magnetic
field:
\begin{eqnarray}
\hspace{-1cm}\textbf{B}(r,z)&=&\{B_r,B_z\} \nonumber \\
                &=&\left\{-\frac{B_1}{2}r-\frac{B_2}{2}rz,B_0+B_1z+\frac{B_2}{2}\left(z^2-\frac{r^2}{2}\right)\right\}+\nonumber \\
   & &\hspace{-.5cm}B_3\left\{\frac{r^3}{16}-\frac{r z^2}{4},\frac{z^3}{6}-\frac{r^2 z}{4}\right\}+
                B_4\left\{\frac{r^3 z}{16}-\frac{r z^3}{12},\frac{z^4}{24}-\frac{r^2 z^2}{8}+\frac{r^4}{64}\right\},
 \label{field}
\end{eqnarray}
with third-order magnitude:
\begin{eqnarray}
 B(r,z)&=&B_0+B_1 z+\frac{1}{2}B_2 z^2+\frac{1}{2}\left(\frac{{B_1}^2}{4B_0}-\frac{B_2}{2}\right)r^2\nonumber\\
       & &+B_3\frac{z^3}{6}+r^2 z \left(-\frac{B_3}{4}+\frac{B_1B_2}{4B_0}-\frac{{B_1}^3}{8{B_0}^2}\right).\label{magnit}
\end{eqnarray}
If, in addition, the coil system is axially-symmetric the  magnitude  to fourth order is:
\begin{eqnarray}
 \hspace{-1.5cm} B(r,z)&=&B_0+\frac{1}{2}B_2 (z^2-r^2/2)+B_4\left(\frac{z^4}{24}+\frac{r^4}{64}\right)+
 \frac{r^2 z^2}{8}\left(-B_4+\frac{{B_2}^2}{B_0}\right).\label{magnit2}
\end{eqnarray}

Let us consider two coils of $N$ turns with radius $a$, separation $s$, carrying currents
$I_1$ and $I_2$.  It is convenient to partition the currents in each coil as a current
$I_{\rm H}$ with the same sense and a current $I_{\rm AH}$ in opposite senses, i.e.
$2I_{\rm H}=I_1 +I_2$, $2I_{\rm AH}=I_1 -I_2$. If one defines $\eta=\mu_0 NI/2,$ and uses
the scaled separation $S = s/a$, the axial magnetic field is thus:
\begin{equation}
B(0,z)=\left(\frac{(\eta_{\rm H}+\eta_{\rm AH})}{a(1+(z/a-S/2)^2)^{3/2}}+\frac{(\eta_{\rm
H}-\eta_{\rm AH})}{a(1+(z/a+S/2)^2)^{3/2}}\right),
\end{equation}
yielding the axial Taylor expansion terms:
\[
 B_0=\frac{2 \eta_{\rm H}}{a\left(1+S^2/4\right)^{\frac{3}{2}}},\;\;
 B_1=\frac{3 \eta_{\rm AH} S}{a^2\left(1+S^2/4\right)^{\frac{5}{2}}},\;\;
 B_2=\frac{6 \eta_{\rm H} \left(S^2 - 1 \right)}{a^3\left(1+S^2/4\right)^{\frac{7}{2}}},
\]
\begin{equation}
 B_3=\frac{15 \eta_{\rm AH} S\left(S^2-3\right)}{a^4\left(1+S^2/4\right)^{\frac{9}{2}}},\;\;
 B_4=\frac{45 \eta_{\rm H} \left(S^4-6S^2+2\right)}{a^5\left(1+S^2/4\right)^{\frac{11}{2}}}.\label{b0b1b2}
\end{equation}
Thus axially symmetric lens systems (equation (\ref{magnit2}): $\eta_{AH}=0,$ $B_{\rm
odd}=0),$ consisting of one or two circular coils have radial and axial strengths:
\begin{equation}
{\omega_{z}}^{2}=-2 {\omega_{r}}^{2}=\frac{3 \mu_0 \mu_B N I (1+{\rm sign}(S)) \left(S^2
- 1 \right)}{2 m a^3\left(1+S^2/4\right)^{7/2}}. \label{omega12}
\end{equation}
The sign function switches between a single and a double coil.
The expressions of equation~(\ref{b0b1b2}) will be utilised in the next section where
various configurations of coils and bars for realising magnetic lenses are considered.

\subsection{Configurations for realising magnetic lenses}

There are six distinct coil and current bar configurations used in this paper for
single-pulse focusing. Figure~\ref{Strategies} displays these six strategies, whilst
figure~\ref{Bfields} shows the associated magnetic field information. Strategies I-III
lead to the axial/radial focusing of equation (\ref{force1D}), whereas strategies IV-VI
yield isotropic 3D focusing (equation (\ref{force})). Strategies I-V deal with
cylindrically symmetric coil arrangements.

\begin{figure}[ht]
\begin{center}
\epsfxsize=\columnwidth \epsfbox{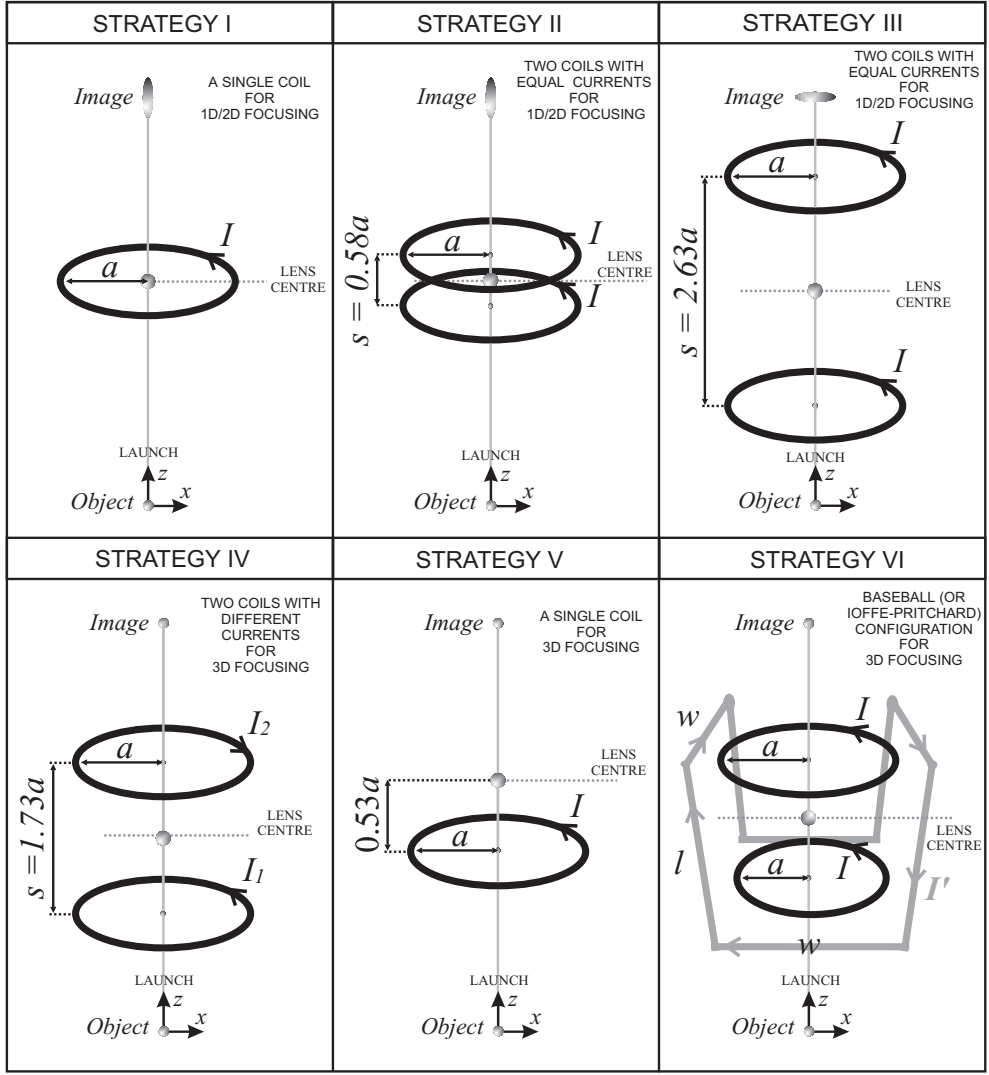}
 \caption{The six different lens
strategies detailed in the text. Strategy I utilises the centre of a single coil;
Strategies II and III use the geometric centre of a pair of coaxial coils (carrying equal
currents in the same sense) with separations of $S$=0.58 and $S$=2.63 coil radii,
respectively; Strategy IV uses the geometric centre of a pair of coaxial coils with
unequal currents and a relative separation $S=\sqrt{3};$ Strategy V uses a single coil
axially offset to $z/a=\pm\sqrt{2/7}$; Strategy VI uses the geometric centre of a
Baseball coil with dimensions $w=l=2a$ combined with a coaxial coil pair with $S=1.$}
\label{Strategies}
\end{center}
\end{figure}

\begin{figure}[ht]
\begin{center}
\epsfxsize=\columnwidth \epsfbox{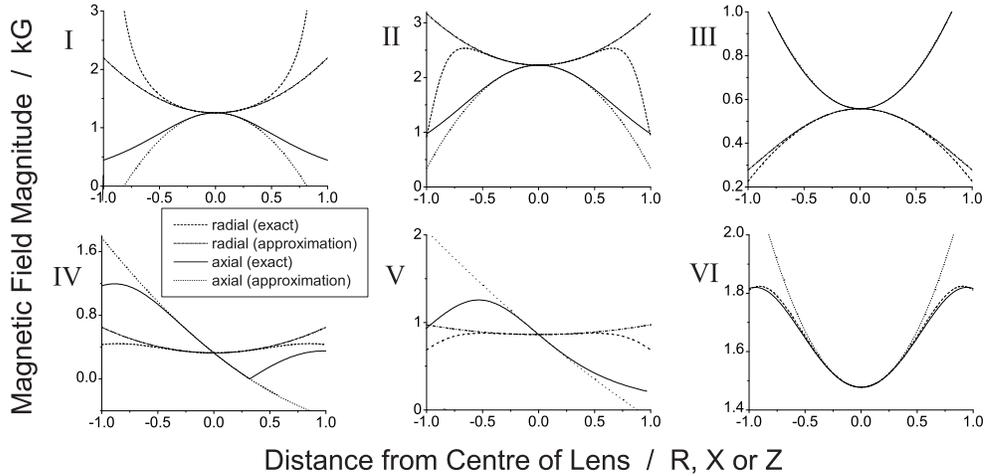}
 \caption{Here we show the respective axial and radial variation of the
magnetic field magnitude associated with the six lens strategies I-VI of
Figure~\ref{Strategies}. $R$, $X$ and $Z$ are the scaled radial, $x$ and axial
displacements from the centre of the lens, respectively.
In Images IV and
V the axial gradients in the magnetic field magnitude alter the centre-of-mass velocity
of an atom cloud but do not lead to lensing. Parabolic approximations to the field
magnitudes are also given.} \label{Bfields}
\end{center}
\end{figure}

\subsubsection{Strategy I: a single coil}

The magnetic field magnitude at the centre of a single coil of radius $a$ with $NI_1$
current turns is characterised by the coefficients in equation~(\ref{b0b1b2}) with $S=0$
and using $I_2=0$ when evaluating $\eta_{\rm H}.$ The radial curvature is positive, thus,
a single coil can be used to radially focus weak-field seeking atoms. However this lens
has an axial curvature that is negative and twice the magnitude.
Figure~\ref{Strategies}~Strategy I shows the geometry of this lens.

\subsubsection{Strategies II, III and IV: a pair of axially-displaced coils}

For a pair of separated coaxial coils, where both coils have equal current with the same
sense, an axially diverging (converging) lens is realised for separations less (greater)
than the coil radius (equation (\ref{omega12})). The lens curvature is zero when the
separation equals the coil radius (the Helmholtz condition for achieving uniform fields).
The axial curvature is a factor of -2 greater than the radial curvature.

Since strategies II and III are axially-symmetric configurations there are no third
order terms. The third
term in equation~(\ref{magnit2}) is zero when $B_4 = 0$, which occurs if $S =
\sqrt{3\pm\sqrt{7}}=0.595$ or 2.38. The fourth term in equation~(\ref{magnit2}) is zero
when $B_4 B_0 = B_2 ^2$, which occurs if $S  = \sqrt{\frac{1}{3}(13\pm\sqrt{145})}=0.565$
or 2.89. The harmonicity of a radial-focusing lens ($S<1$ Strategy II) is thus optimised
if the relative coil separation is $S=0.58,$ whereas the harmonicity of an axial-focusing
lens ($S>1$ Strategy III) is thus optimised if the relative coil separation is $S=2.63.$
Figure~\ref{Strategies}~II and III show the geometry of these optimised lenses.

Figure~\ref{Strategies}~Strategy IV shows the geometry of an isotropic 3D lens formed
from two coils carrying different currents. From Equation (\ref{magnit}) we see that an
isotropic 3D lens is formed when ${B_1}^2=6B_2 B_0,$ and this can be re-expressed using
equation (\ref{b0b1b2}) as:
\begin{equation}
\eta_{\rm AH}=\pm\eta_{\rm H} \sqrt{8(1-1/S^2)}.\label{isocurrent}
\end{equation}
From equations~(\ref{magnit},\ref{b0b1b2}) we see that some of the third-order terms in
the magnetic field magnitude are removed by setting $S=\sqrt{3}$ and thus $B_3=0.$ This
means that the coils carry currents of $I_1$ and $I_2=-0.396 I_1$ respectively. Note that
the only way to remove all third-order terms is to have $S=\sqrt{3}$ and set
${B_1}^2=2B_2 B_0$ (i.e. $\eta_{\rm AH}=\pm\eta_{\rm H} \sqrt{8/3(1-1/S^2)}$~) which
corresponds to a purely axial lens. Purely radial lenses are achieved when $B_2=0,$ (i.e.
$S=1)$ for non-zero $\eta_{\rm AH}$ and $\eta_{\rm H}.$

\subsubsection{Strategy V: an axially offset single coil}\label{isotropicII}

At the centre of a single coil the radial and axial curvatures have opposite signs
(Strategy I), however, the radial and axial curvatures have different $z-$dependence,
(Figure~\ref{Strategies}~Strategy V). At the two axial locations $z=\pm \sqrt{2/7}a$ the
curvatures are equal in magnitude and both positive. Therefore an impulse applied to a
cloud whose centre-of-mass is at either of these positions will lead to isotropic 3D
focusing.

\subsubsection{Strategy VI: Ioffe-Pritchard configuration}

Ioffe-Pritchard (IP) traps are used extensively for atom trapping and are similar to the
Ioffe configuration utilised in plasma confinement~\cite{IPrefs}. One of the simplest
forms of this trap is with a pair of coaxial coils to generate both a bias field, $B_0$,
and an axial curvature $B_2$.  Four current bars run parallel to the $z$-axis, with each
bar running through the corner of a square (with side length $w=W a)$ centred on the
axis. The magnitude of the current in each bar is equal, but neighbouring bars have the
opposite current sense.

This configuration is not strictly radially symmetric, although it is an excellent
approximation for small displacements from the axis. The four current bars give rise to a
transverse magnetic field of the form $\{B_x,B_y\} = B_{1}' \{x,-y\}$ close to the axis.
Here, $B_{1}' = 4 \mu_0 I'/(\pi w^2)$, where $I'$ is the magnitude of the current in a
bar, and the prime distinguishes it from the axial gradient defined previously. The shape
of the magnetic field magnitude contours depend on the geometry of the trap and the ratio
of the coil current to the bar current. Such traps are extensively used in Bose-Einstein
condensation experiments, where an additional circular coil pair is used primarily to
lower the bias field $B_0,$ yielding an anisotropic harmonic-oscillator potential, with
tight radial confinement $B_{2r}={B_{1}'}^{2}/B_0-B_2/2$, and weaker axial confinement
$B_{2z}=B_2$. For the purposes of magnetic focusing an isotropic IP trap is required.
This is achieved if the following relationship holds among the lens parameters:
\begin{equation}
\frac{I'}{NI}= 3 \pi W^2\sqrt{\frac{S^2-1}{(2+S^2/2)^5}}.
\end{equation}

In BEC experiments variants of the IP trap described above are used, since
infinitely-long current bars are not realisable. An elegant winding pattern is the
baseball geometry. A baseball coil has an axial curvature  in addition to a radial
gradient, and these quantities can no longer be independently varied.
Figure~\ref{Strategies}~Strategy VI shows the cuboidal baseball geometry, where the sides
have lengths $w=W a$, $w=W a$ and $l=L a$, ($l$ is along the $z$-axis). The bars carry a
current $I'$. It is impossible to realise a 3D isotropic lens solely by adjusting the
aspect ratio of the coils $w/l,$ and for this reason we add a coaxial pair of coils
carrying equal currents in the same sense (if $\eta_{\rm AH}\neq 0$ then the equality of
the $x$ and $y$ curvatures is broken). The magnetic field magnitude has a third-order
Taylor expansion:
\begin{equation}
 \!\!\!\!\!\!\!\!\!\!\!\!\!\!\!\!\!\!\!\!\!\!\!\!
 B=B_{\rm f0}+\left(\frac{{B_{1}'}^{2}}{2B_{\rm f0}}-\frac{B_{\rm f2}}{4}\right)
(x^2+y^2)+\frac{B_{\rm f2}}{2} z^2+\left(B_3'-\frac{B_{\rm f2} B_{1}'}{2 B_{\rm
f0}}\right)(y^2-x^2)z,\label{baseTay}
\end{equation}
where dashed terms indicate contributions solely from the baseball coil, and the
subscript f is used when one must add together the $B_0,$ $B_2$ Taylor contributions from
the circular coils (equation (\ref{b0b1b2})) and $B_0',$ $B_2'$ from the following
baseball coil Taylor terms:
\begin{eqnarray}
 \!\!\!\!\!\!\!\!\!\!\!\!\!\!\!\!\!\!\!\!\!\!\!\!
 B_0'=\frac{\frac{4\mu_0 I'}{\pi}W^2}{a\left( L^2 + W^2 \right) {\sqrt{L^2 + 2\,W^2}}},\,\,
 B_1'=\frac{\frac{4\mu_0 I'}{\pi}\left( L^5 + 3L^3 W^2 + 4L W^4 \right)}{a^2 W^2 {\left( L^2 + W^2 \right) }^2 {\sqrt{L^2 +
 2 W^2}}}\nonumber \\
 B_2'=\frac{\frac{16\mu_0 I'}{\pi}\left( 6 L^6 W^2 + 18 L^4 W^4 + 11 L^2 W^6 - 5 W^8 \right)}
  {a^3\,{\left( L^2 + W^2 \right) }^3\,{\left( L^2 + 2 W^2 \right)
  }^{\frac{5}{2}}},\nonumber \\
 B_3'=\frac{\frac{48\mu_0 I'}{\pi}\left( -5 L^7 W^2 - 10 L^5 W^4 + 11 L^3 W^6 + 24 L W^8 \right)}
  {a^4 {\left( L^2 + W^2 \right) }^4 {\left( L^2 + 2 W^2 \right) }^{\frac{5}{2}}}.
\label{basebob1b2}
\end{eqnarray}

The simplest isotropic lens to calculate is when we use circular Helmholtz coils -- i.e.
$S=1,$ $B_2=0$ and the only contribution of the circular coils in equation
(\ref{baseTay}) is the axial constant field $B_0,$ which we tune to enable lens isotropy.
The finite-length current bars lead to a non-trivial relationship among the parameters
for realising an isotropic lens by adjusting the circular coil bias-field and curvature.
For the purpose of our calculations we use the parameters $W=W=L=2$ for a cubic baseball,
with $S=1$ and a relative circular coil current of $\frac{NI}{I'}=0.154.$

\subsection{The harmonicity of the magnetic lenses}

The previous subsection was a description of the way to achieve a radially converging lens
with a single coil or a pair of coils; an axially converging lens with a pair of
displaced coils; and three geometries for achieving an isotropic 3D lens. Aberrations are
caused by departures of the actual potential from the ideal harmonic potential. To
measure the degree of departure from harmonicity of a lens, as a function of the distance
from the lens' centre, we use:
\begin {equation}
\epsilon=\frac{|{\bf a}_{\rm F}-{\bf a}_{\rm H}|}{|{\bf a}_{\rm H}-{\bf a}_{\rm 0}|},
\label{harmonicity}
\end{equation}
where ${\bf a}_{\rm H}$ is the harmonic fit, (equation (\ref{force1D}) or (\ref{force})),
to the full Biot-Savart lens acceleration ${\bf a}_{\rm F}.$ The departure from
harmonicity is a function of position, and  the cylindrical co-ordinates $R$ and $Z$ are
used to plot $\epsilon(R,Z)$ for different coil systems. Note the scaled co-ordinates $R$
and $Z$ are normalised to measure length in units of the coil radius,  $R=r/a$ and
$Z=z/a$. Figure (\ref{Fig2}) shows the departure from harmonicity for the six different
focusing strategies.

\begin{figure}[ht]
\begin{center}
\epsfxsize=\columnwidth \epsfbox{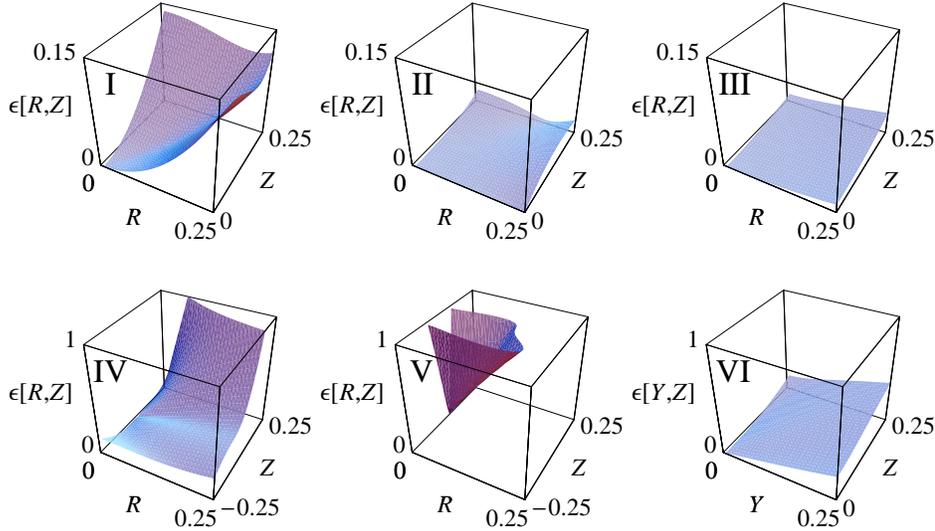}
 \vspace{-1.3cm}\caption{The departure from harmonicity $\epsilon(R, Z)$ of the six different lens
strategies of Fig.~\ref{Strategies} is considered for displacements of up to 0.25 coil
radii. Averaged over a sphere of 0.25 times the coil radius, the departures from
harmonicity for the six configurations are: 0.058, 0.007, 0.009, 0.385, 1.347 and 0.148
respectively. }\label{Fig2}
\end{center}
\end{figure}

\subsubsection{Strategies I-III: axial/radial focusing}
Figures~\ref{Fig2}~I-III show the spatial variation of the departure from harmonicity of
Strategies I-III. For Strategy I, II and III the departure from harmonicity averaged over
a sphere of radius $0.25\,a$ is 0.058, 0.007 and 0.009 respectively. The radially
converging lens of Strategy II, and the axially converging lens of Strategy III are
almost an order of magnitude more harmonic than the single coil radially converging lens
of Strategy I.

\subsubsection{Strategies IV-VI: isotropic 3D focusing}
The isotropic 3D lens of Strategy IV rapidly becomes anharmonic as one moves axially away
from the lens centre because of the axial magnetic field zero illustrated in
Figure~\ref{Bfields}~IV. In both Strategy IV and V the lack of axial symmetry means that
there are first-order terms in the field magnitude, resulting in a gradient which affects
the centre-of-mass motion of the atomic cloud.

The main problem with Strategy V is that although both the axial and radial curvatures
are equal at the lens centre, they vary rapidly with position. The lens-centre curvatures
are also 8.4 times weaker than the axial curvature at the centre of a single coil,
leading to longer duration magnetic lensing impulses. The cloud of atoms therefore experiences
the lens' axial anharmonicities for a greater period, again compromising the
quality of the focus.

Figures~\ref{Fig2}~IV-VI show the spatial dependence of the departure from harmonicity of
Strategies IV-VI. For Strategy IV, V and VI the departure from harmonicity averaged over
a sphere of radius $0.25\,a$ is 1.347, 0.385 and 0.148 respectively. The IP lens of
Strategy VI is thus significantly more harmonic than the two coil lens of Strategy IV, which is
in turn considerably better than the axially offset single coil lens of Strategy V.

\section{Imaging and minimum cloud size}\label{imagesec}
The separable equations of motion for a lens that is harmonic in 3D
(equations (\ref{force1D}),(\ref{force})) allow us to consider
motion in each cartesian dimension as a separate simple harmonic
equation. It is useful to employ the $\mathcal{ABCD}$-matrix
formulation used widely in geometrical optics. The position and
velocity of an atom along a given Cartesian axis, say $x$, is
written as a 2-component vector, and the final and initial vectors
are related via the equation:
\begin{equation}
\left( \begin{array}{c} x_{\rm f}  \\
          v_{x_{\rm f}} \end{array} \right)=\left( \begin{array}{cc}
\mathcal{A} & \mathcal{B}  \\
\mathcal{C}  & \mathcal{D}
\end{array} \right)
\left( \begin{array}{c} x_{\rm i}  \\
          v_{x_{\rm i}} \end{array} \right).
\label{eq:ABCD}
\end{equation}
To simplify the effects of gravity we perform the calculations in a
free-falling frame of reference.  In this frame the free evolution of the
cloud is an isotropic expansion, described by an $\mathcal{ABCD}$ matrix.
The influence of a converging or diverging magnetic lens
can also be described by  $\mathcal{ABCD}$ matrices, as outlined in \cite{Gor}:
\begin{equation}
\!\!\!\!\!\!\!\!\!\!\!\!\!\!\!\!\!\!\!\!\!\!\!\!\!\!\!\!\!\!\!\!
M_1(t)\!=\!\left(\!\!\begin{array}{cc}
1 \!\!& t \\
0 \!\!& 1
\end{array}\!\!\right),
M_2\!=\!\left(\!\!\begin{array}{cc}
\cos \omega \tau\!\!\!\! & \frac{1}{\omega}\sin \omega \tau \\
-\omega \sin \omega \tau\!\!\!\! & \cos \omega \tau
\end{array}\!\!\right),
M_3\!=\!\left(\!\!\begin{array}{cc}
\cosh \omega \tau \!\!\!\!& \frac{1}{\omega} \sinh \omega \tau \\
\omega \sinh \omega \tau \!\!\!\!& \cosh \omega \tau
\end{array}\!\!\right).
\label{mats}
\end{equation}
Matrix $M_1$ is the translation matrix for a duration $t$; $M_2$ is
the matrix for a converging lens of strength $\omega$ applied for a
duration $\tau$; $M_3$ is the matrix for a diverging lens of
strength $\omega$ applied for a duration $\tau$. It is interesting
to note that the sinusoidal (exponential) path taken by atoms inside
a converging (diverging) magnetic lens is in stark contrast to the
linear propagation of light rays in an optical lens.

By multiplying these matrices together, we arrive at the final
$\mathcal{ABCD}$ system matrix. An image (i.e. a one-to-one map of
position between the initial and final cloud) is formed if the
condition $\mathcal{B}=0$ is maintained. In this case the spatial
magnification $\mathcal{A}$ is the inverse of the velocity
magnification $\mathcal{D}.$ This spatial compression and
concomitant velocity spread is a manifestation of Liouville's
theorem. The theorem states that phase-space density is conserved in
a Hamiltonian system. Time-dependent Stern-Gerlach forces satisfy
the criteria for Liouville's theorem to be valid \cite{Ketterle92}.
The cloud extent along $x$ in a given plane is given by:
\begin{equation}
\sigma_{x_{f}}^{2} = (\mathcal{A}\sigma_{x_{i}})^2 +(\mathcal{B}
\sigma_{v_{x_{i}}})^2, \label{size}
\end{equation}
where $\sigma_{x_i}$ is the initial position standard deviation and
$\sigma_{v_{x_i}}$ is the initial velocity standard deviation. An
{\it image} is formed for the condition $\mathcal{B}=0$, but the
{\it smallest cloud size} does not necessarily occur in the same
plane. For a single lens system, the minimum cloud size occurs very close
to the imaging plane. For multi-lens systems, the
image plane and the minimum cloud size do not necessarily
correspond.

\subsection{Imaging solutions}\label{imagesol}

We note here that the `thick' converging lens $M_2$ is identical to a thin lens of
strength $\mathcal{C}(\omega,\tau)=-\omega \sin(\omega \tau)$ (i.e.~the original
$\mathcal{C}$ entry of the $M_2$ $\mathcal{ABCD}$ matrix), pre- and post-multiplied by a
translation matrix with duration $\tau'/2$:
\begin{equation}
M_2=\left( \begin{array}{cc}
1 & \tau'/2 \\
0 & 1
\end{array} \right)
\left( \begin{array}{cc}
1 & 0 \\
\mathcal{C} & 1
\end{array} \right)
\left( \begin{array}{cc}
1 & \tau'/2 \\
0 & 1
\end{array} \right).
\label{thincon}
\end{equation}
The pulse width $\tau'$ is defined as
\begin{equation}
\tau'(\omega,\tau)=\frac{2}{\omega}\tan\frac{\omega \tau}{2}, \label{tauprime}
\end{equation}
 and  the notation of primes is used to denote times in the
`thin' lens representation. This means that we can use many of the simplicities of `thin'
lens optics, even if we are in fact dealing with the more accurate `thick' lensing
behaviour. The effective `thin lens' duration of the pulse $\tau'$ differs from the
actual pulse duration $\tau,$ but otherwise the treatments are identical. In the limit of
a short, strong pulse $\omega \tau\rightarrow 0,$ we find that $\tau'\rightarrow\tau.$ If
we wish to consider the diverging lens $M_3,$ we merely make the transformation
$\omega\rightarrow i\omega$ in equation (\ref{thincon}) --
i.e.~$\mathcal{C}=\omega\sinh(\omega \tau)$ and $\tau'=\frac{2}{\omega}\tanh\frac{\omega
\tau}{2}.$

We can model a single lens system by having a translation of $M_1(t'_1),$ where
$t_1'=t_1+\tau'/2,$ followed by a thin lens of strength $\mathcal{C},$ followed by a
translation of $M_1(t'_2),$ where $t_2'=t_2+\tau'/2.$ The physical duration of the
focusing is $T=t_1+t_2+\tau,$ however the thin lens system has a total time
$T'=t_1+t_2+\tau'=T-\tau+\tau'.$ For a single lens system, the condition $\mathcal{B}=0$
is met for the system $\mathcal{ABCD}$ matrix if we have:
\begin{equation}
\mathcal{C} T' = \frac{1}{\lambda (\lambda-1)}, \label{Bzero}
\end{equation}
where
\begin{equation}
\lambda=\frac{t_1'}{T'},
\end{equation}
 yielding a magnification
$(\lambda-1)/\lambda.$ This formalism provides a useful way of
designing a lens system and investigating its focusing properties.
For a converging lens, equation (\ref{Bzero}) becomes:
\begin{equation}
\omega T' \sin \omega \tau  = \frac{1}{\lambda (1-\lambda)}.
\label{Bzerocon}
\end{equation}
We consider an experimental situation where we fix the total time $T=212\,\rm{ms}$ and
the maximum Amp-turns at $NI=10,000\,$Amps. The geometry of the lens then fixes the
maximum strength of the converging lens via ${\omega_r}^2=\mu_\mathrm{B}B_2/m$. We can
now solve to find an analytic result for $\lambda(\omega,\tau),$ which is illustrated in
figure~\ref{maxfoc} for a single coil lens of radius 5~cm
($\omega_r=69.6\,\rm{rad}\,\rm{s}^{-1}$ from equation (\ref{omega12})). The
$\lambda(\omega,\tau)$ parameter is maximised (and the magnification
$(1-\lambda)/\lambda$ is minimised) when:
\begin{equation}\label{cot}
1-\omega (T-\tau) \cot \omega \tau=0,
\end{equation}
which has the solution $\lambda=0.929$ at $\tau=23.7\,\rm{ms}.$ This corresponds to a
reduction in the atomic cloud size by a factor of $-13.1$ This is achieved when the pulse
duration $\tau$ is from time $t=T-\tau$ to $t=T,$ i.e.~the lens pulse ends at the time of
focus. Such focusing in three dimensions would increase the cloud density by more than 3
orders of magnitude. For a lens placed later in time, the magnetic pulse would not have
finished at the predicted focal time $T$, resulting in an increase in cloud size at time
$T$.

\begin{figure}[ht]
\begin{center}
\epsfxsize=\columnwidth \epsfbox{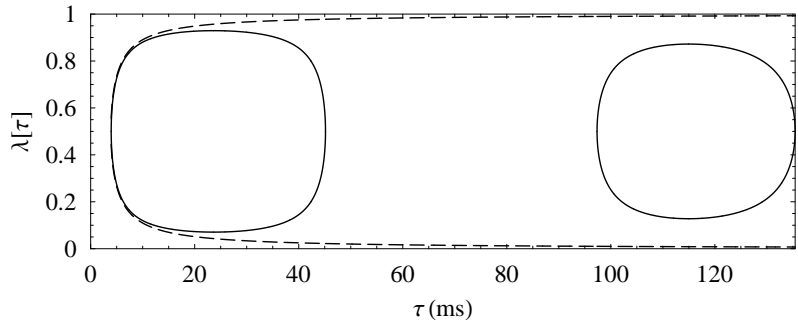}
 \caption{By fixing the strength of a $5\,$cm Strategy I radially converging lens,
 $\omega_r=69.6 \rm{rad}\,\rm{s}^{-1},$ and total
experimental focusing time $T=212\,\rm{ms},$ we can vary the lens pulse width, $\tau$,
and find the focusing parameter $\lambda(\tau)$ via equation (\ref{Bzerocon}). The
impulse $\tau$ has a minimum of $3.9\,\rm{ms}$ at $\lambda=1/2$, and $\lambda$ is
symmetric about this point. Also shown (dashed) is the result obtained if one makes the
strong, short pulse approximation $\omega \tau\rightarrow 0,$ leading to the
simplification $\sin\omega \tau \approx \omega \tau$ in equation (\ref{Bzerocon}) -
resulting unsurprisingly in a divergence for large pulse durations.}\label{maxfoc}
\end{center}
\end{figure}

The above analysis would seem to suggest that the optimum strategy for achieving the
smallest cloud size would be to construct a lens with a short, strong pulse
$\omega\tau\rightarrow0$, and use the latest possible pulse time $\lambda\rightarrow1$.
However, experimental constraints and lens aberrations alter the above conclusion.

\section{Investigating various coil configurations for pulsed focusing}

\subsection{Methodology}

The $\mathcal{ABCD}$ matrix formalism outlined above is a useful starting point for
studying pulsed magnetic focusing. However, this formalism ignores magnetic aberrations
arising due to the departure of the real potential from an ideal parabolic spatial
dependence (figure~\ref{Fig2}). For a non-parabolic potential the change in position and
velocity which occur during lensing must be calculated numerically. Here, for the first
time to our knowledge, we test the `perfect' atomic lens approximations by performing
numerical focusing simulations. The cloud and its motion are treated classically, and for
the atomic densities encountered in the expanding cloud, the collision rate is
negligible. The atoms travel on ballistic trajectories, except when a magnetic impulse is
applied, in which case the full Stern-Gerlach force is included in the numerical
integration. For the sake of definiteness we chose to investigate the focusing of
$^{85}$Rb atoms in an atomic fountain launched vertically through a height of $22\,$cm
(which corresponds to a flight-time of $212\,$ms), a height which is of interest
experimentally. The atoms come to rest at the apex of their trajectories where they could
be used for further experiments or loaded into a dipole trap.  The effects of gravity
were included, but these effects on the quality of focus were found to be negligible for
the parameters used in these simulations.

The approach adopted was a numerical simulation, in which the trajectories of typically
500  atoms are followed.  The initial velocity and position probability distributions are
isotropically Gaussian for each Cartesian direction. The standard deviation of position
was chosen to be a value typical of experiments at $0.4\,$mm \cite{Hinds}, and the
velocity distribution corresponding to a typical launch temperature obtained with Rb in
moving molasses, namely $T=20\,\mu$K. These simulations facilitate the calculation of
statistically relevant quantities, such as the standard deviation of the time-dependent
size of the atomic cloud. In all the simulations the maximum current value in any coil
was limited to $10,000\,$A.

\subsection{Strategies I-III: axial/radial focusing}

To illustrate our methodology we discuss the focusing properties of a single, circular,
current-carrying coil (Strategy~I). Figure~\ref{single}~(a) shows the evolution of the
$x$ component of a launched cloud of 500 atoms subject to a radially converging lens
constructed from a single $5\,$cm radius coil. The impulse is applied half-way in time,
and the length of the impulse is chosen to reverse the transverse velocity, as can be
seen from the change in sign of the gradient after $t=106\,$ms.  For this case, the
$\mathcal{ABCD}$ matrix predicts a  radial focus with magnification $-1,$ which is in
excellent agreement with the numerical simulation using a parabolic lens (using the
strength of equation~(\ref{omega12})). The vertical line at $t=212\,$ms corresponds to
the imaging time. For real coils, it is seen that the focussed cloud image is
significantly larger than the initial cloud. The aberration worsens as the coil radius
decreases. Note that although focusing is seen in the radial direction, defocusing is
seen in the axial direction due to the opposite sign of the magnetic field curvature.

\begin{figure}[ht]
\begin{center}
\epsfxsize=\columnwidth \epsfbox{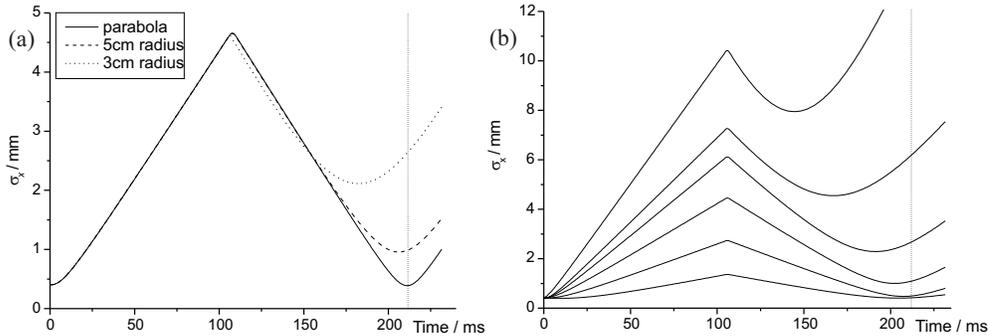}
 \caption{\label{single} (a) Simulation of 500 atoms going through a $10,000\,$Amp-turn
strategy I lens.  The solid, dashed and dotted lines correspond to a parabolic
approximation lens, a $5\,$cm radius lens, and a $3\,$cm radius lens respectively. The
duration and timing of the impulse is chosen using the $\mathcal{ABCD}$ matrix formalism,
bringing the atoms to a focus $22\,$cm above their launch height. (b) A shell plot of
1000 simulated atoms passing through a  $3\,$cm radius strategy I lens.
The distances from the
coil centre are 0-10\% of the coil radius, 10-20\%, through to 50-60\%. Atoms further
from the centre are not focussed as well, and the focusing occurs at earlier times.  Both
of these factors degrade the image quality and size.}
\end{center}
\end{figure}

Figure~\ref{single}~(b) contains an analysis of the cloud in terms of shells of different
radii measured from the centre of the coil; atoms further from the centre are not
focussed as tightly, and also focus earlier in time. As the ratio between cloud extent
and coil radius decreases, the departure of the field from the parabolic approximation
becomes less significant. Therefore one method to reduce the aberrations experienced by
the atoms is to increase the coil radius, or decrease the atomic cloud temperature.

We turn our attention to obtaining the minimum cloud size, equation~(\ref{size}), by
investigating the effect of the $\lambda$ parameter. As discussed in
section~\ref{imagesol}, a thin parabolic lens produces the smallest cloud size when
$\lambda \rightarrow 1$, i.e. the pulse is applied as late in time as possible. This is a
manifestation of Liouville's theorem - a more compact spatial extent can be generated at
the expense of a larger velocity spread.

Figure~\ref{strat}~(a) shows simulations of radial focusing. The radial cloud expansion
factor, $\sigma_{x}/\sigma_{x_i}$, is plotted as a function of the imaging parameter
$\lambda$. It can be seen that the smallest cloud size for a parabolic lens occurs when
one waits as long as possible before focusing, i.e.~$\lambda$ is as close as possible to
1 (limited by the solution of equation~(\ref{cot})). This means that the magnetic impulse
ends at the desired focusing time. For parabolic fits to the 3 and $5\,$cm radius single
coils (Strategy I) the maximum values of $\lambda$ are 0.968 and 0.929, corresponding to
negative magnifications, $M=(1-\lambda)/\lambda$, of $1/30.1$ and $1/13.1$ respectively.

\begin{figure}[ht]
\begin{center}
 \epsfxsize=\columnwidth \epsfbox{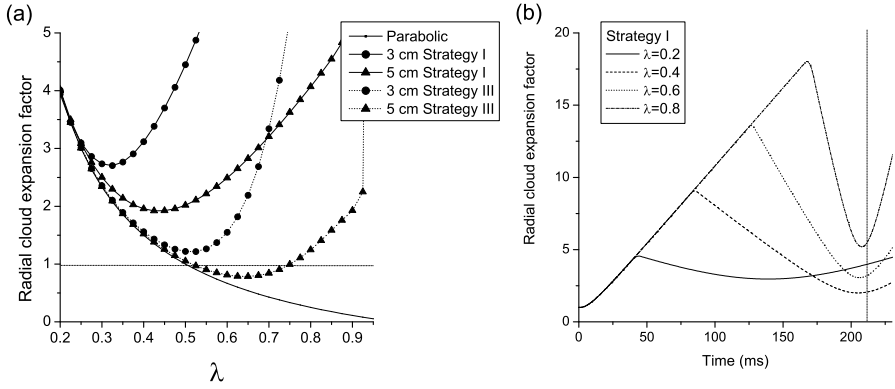}
 \vspace{-1cm}\caption{\label{strat}(a) The radial
expansion factor, at $T=212\,$ms, for radially converging lenses is plotted against
$\lambda$. The solid line without symbols shows the parabolic lens solution, the solid
(dotted) lines with symbols shows the result of atoms passing through Strategy I (II)
lenses. Circular (square) symbols are used for the $3\,$cm ($5\,$cm) radius lenses. The
Strategy II lens is more parabolic, allowing a smaller radial cloud size to be achieved:
a cloud image 0.75 times the original size occurs for $\lambda=0.65$ using a $5\,$cm lens.
(b) For a $5\,$cm Strategy I lens, the radial expansion factor is plotted against time
for $\lambda$ values $0.2,\,0.4,\,0.6$ and $0.8.$}\end{center}
\end{figure}

Along with the parabolic case, figure~\ref{strat}~(a) shows numerical simulations for 500
atoms passing through 3 and $5\,$cm radius coils, for both Strategies I (single coil) and
II (two coils). As expected the $5\,$cm lens better approximates a parabola. Compared to
the $\mathcal{ABCD}$ matrix result there is a marked difference in the behaviour of the
minimum radially-focussed cloud-size for fields from real coils -- the value of $\lambda$
at which the $\mathcal{ABCD}$ minimum is obtained is dominated by aberrations in the
magnetic field. The $\mathcal{ABCD}$ matrix approach does not provide an adequate
description of pulsed magnetic focusing when one considers the entire atomic cloud.

In figure~\ref{strat}~(b) the radial expansion factor of a $5\,$cm single coil (Strategy
I) is plotted against time for values of $\lambda$ varying from $0.2$ to $0.8$.

The easiest way to reduce aberrations appears to be the use of a very large coil radius.
Unfortunately the curvature of the field decreases with the cube of the coil radius
(equation~(\ref{omega12})), which necessitates longer pulse durations for larger radius
coils. This increase in pulse duration reduces the maximum value of $\lambda$ that can be
used and therefore also limits the minimum cloud size. The aberrations can only be
further reduced by increasing the current-turns, something which has experimental
limitations.

The aberrations associated with a real coil dramatically affect the strategy for
achieving a radially compact cloud.  The impulse has to be applied significantly earlier
than an $\mathcal{ABCD}$-matrix analysis would suggest. However, for
experimentally-realistic parameters it is seen that it is possible to achieve a final
radial cloud size that is smaller than the initial size. Strategy II (two coils of radius
$a$ carrying the same current $I,$ separated by $S=0.58),$ clearly has a better
performance than Strategy I (single coil), as seen in figure~\ref{strat}~(a). Most
importantly, in the 5~cm case, the radial extent of the final cloud is $0.75$ times
 the original radial extent of the  cloud.

Strategy III produces an axially converging/radially diverging lens with a high level of
harmonicity (figure~\ref{Fig2}) comparable to that of Strategy II. We omit our results
for axial focusing in this paper, as it will be revisited in the context of
alternate-gradient focusing in a future publication~\cite{matt2}.

\subsection{Strategies IV-VI: isotropic 3D focusing}

Isotropic 3D focusing can be achieved using two coils with differing currents (Strategy
IV). For realistic experimental parameters, the numerical simulations showed that the
aberrations in the lens smeared out any focusing. However, for unrealistically large lens
radii and large currents (e.g. $15\,$cm and $200,000\,$Amp-turns) it is possible to
achieve 3D focusing.

As discussed in section~\ref{isotropicII}, a single coil can be made to have isotropic
curvature (Strategy V). At $z=\pm\sqrt{2/7}a$, the axial and radial curvatures are equal,
and the gradient of the field is non-zero. A numerical simulation was performed for a
launched cloud, with an impulse applied when the cloud's centre of mass reached a
distance $z=+\sqrt{2/7}a$ from the centre of a single coil.  Due to the large departure
from harmonicity for the experimentally realistic parameters we used, aberrations
dominated and focusing was not observed.

The baseball lens (Strategy VI) yielded the best isotropic 3D lens.
Figure~\ref{ioffe}~(a) shows the temporal evolution of the volume expansion factor,
($\sigma_{x}\sigma_{y}\sigma_{z}/\sigma_{x_i}\sigma_{y_i}\sigma_{z_i}$), for a launched
cloud subject to a focusing pulse from a baseball lens. Five different values of
$\lambda$ are depicted, from 0.3 to 0.7 in steps of 0.1.  The bias coils have radii of
$a=5\,$cm, separation $S=15$, and current $NI=3082\,$A; the baseball has sides of length $W=L=2$ and carries a
$10,000\,$A current. The minimum cloud size is obtained when $\lambda = 0.3$, and
represents a $31.2$ increase in cloud volume at the focal time, $T=212\,$ms.  This is to
be contrasted with the $13,000$ increase in cloud volume if no magnetic lens were used.
It is interesting to note that for the parameters we have simulated, the results of a
`pure' Ioffe-Pritchard lens ($W=2$, $L\rightarrow\infty$ and $S=2$) are almost
identical to the baseball coil. Surprisingly the baseball lens performs better, and
produces smaller cloud sizes at the focus.

\begin{figure}[ht]
\begin{center}
\vspace{-0.5cm} \epsfxsize=\columnwidth \epsfbox{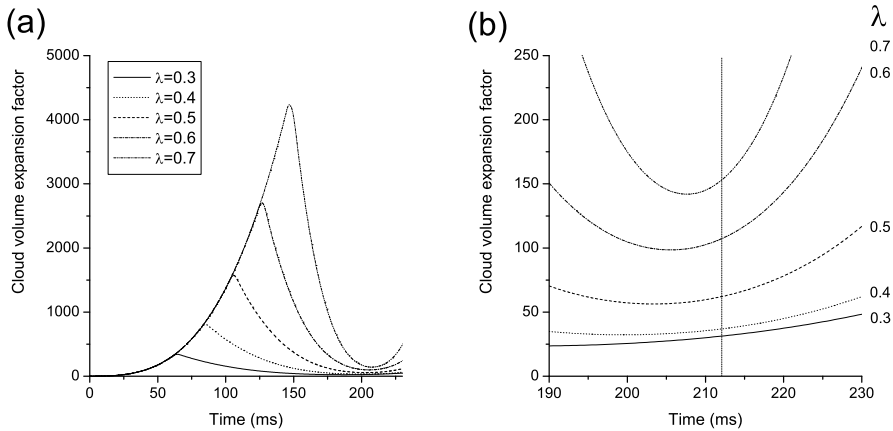} \vspace{-1cm}
\caption{\label{ioffe}(a) A simulation of 500 atoms sent through an isotropic
 baseball coil lens (Strategy VI). The ratio of the cloud volume to the initial volume is
plotted as a function of time for values of $\lambda$ ranging from 0.3 to 0.7. (b) A
close-up of the simulations near the $t=T$ imaging time (vertical line).}\end{center}
\end{figure}

\section{Discussion and Conclusion}
\label{sec:5}

We have highlighted the limiting size of a focussed launched cold cloud of weak-field
seeking atoms for various pulsed magnetic focusing techniques.  The $\mathcal{ABCD}$
matrix formalism is convenient for giving an estimate as to the parameters needed for
magnetic focusing, but does not contain the departure of the potential experienced by
atoms from a perfect parabolic dependence for fields produced by real coils (and bars).
In this work we have shown how important it is to consider these aberrations as they
drastically alter the results. We have identified the origin of these aberrations, and
described techniques for minimising them.

We have discussed single coil lenses (Strategy I), as well as five novel atomic lenses
strategies (II-VI), and tested their aberrations both analytically and numerically. For
all of the Strategies we found that our analytic results for aberrations (section 3) tied
in well with the numerical simulations of section 4. It was demonstrated in Strategies II
and III that a `doublet' radially (axially) focusing lens formed from two coils with
relative separation $S=0.58$ (2.63) provided much tighter focusing than the single-coil
lens of Strategy I. Amongst the isotropic 3D lenses we found that the baseball
lens (Strategy VI) was superior to the two coil lens of Strategy IV, which was in turn
considerably better than the axially offset single coil lens of Strategy V. Of the
single-impulse lenses, the baseball lens offers the best possibilities for
isotropically focusing a cloud of weak-field-seeking atoms in 3D. Experiments to test
these predictions are underway in our laboratory.

It should be noted that in section 4 we have used the rms radius of a cloud of
atoms to measure how tightly the \textit{entire} atomic cloud is focused. By only
considering a low-velocity fraction of the atomic distribution, even fractions as large
as $50\%,$ it is possibly to  reduce drastically the rms focal spot size, increasing the
atomic density by orders of magnitude. We will address this complex issue in more detail
in our future publication dealing with alternate-gradient focusing \cite{matt2}.

\ack This work is supported by EPSRC, the UKCAN network and Durham University. We thank
Charles Adams and Simon Cornish for fruitful discussions.

\end{document}